\documentclass[conference]{IEEEtran}
\IEEEoverridecommandlockouts
\usepackage{cite}
\usepackage{amsmath,amssymb,amsfonts}
\usepackage[linesnumbered,ruled,vlined]{algorithm2e}
\usepackage{algorithmic}
\usepackage{graphicx}
\usepackage{textcomp}
\usepackage{xcolor}
\usepackage{listings}
\usepackage{comment}
\def\BibTeX{{\rm B\kern-.05em{\sc i\kern-.025em b}\kern-.08em
    T\kern-.1667em\lower.7ex\hbox{E}\kern-.125emX}}

\begin{document}

\title{
Unlocking Advanced Graph Machine Learning Insights through Knowledge Completion on Neo4j Graph Database\\
}

\author{
\IEEEauthorblockN{Rosario Napoli}
\IEEEauthorblockA{\textit{University of Messina}\\
Messina, Italy \\
0009-0006-2760-9889} 
\and
\IEEEauthorblockN{Antonio Celesti}
\IEEEauthorblockA{\textit{University of Messina}\\
Messina, Italy \\
0000-0001-9003-6194} 
\and
\IEEEauthorblockN{Massimo Villari}
\IEEEauthorblockA{\textit{University of Messina}\\
Messina, Italy \\
0000-0001-9457-0677} 
\and
\IEEEauthorblockN{Maria Fazio}
\IEEEauthorblockA{\textit{University of Messina}\\
Messina, Italy \\
0000-0003-3574-1848} 
}

\maketitle

\begin{abstract}
Graph Machine Learning (GML) with Graph Databases (GDBs) has gained significant relevance in recent years, due to its ability to handle complex interconnected data and apply ML techniques using Graph Data Science (GDS).
However, a critical gap exists in the current way GDB-GML applications analyze data, especially in terms of Knowledge Completion (KC) in Knowledge Graphs (KGs). In particular, current architectures ignore KC, working on datasets that appear incomplete or fragmented, despite they actually contain valuable hidden knowledge. This limitation may cause wrong interpretations when these data are used as input for GML models.

This paper proposes an innovative architecture that integrates a KC phase into GDB-GML applications, demonstrating how revealing hidden knowledge can heavily impact datasets' behavior and metrics. For this purpose, we introduce scalable transitive relationships,  which are links that propagate information over the network and modelled by a decay function, allowing a deterministic knowledge flows across multiple nodes.

Experimental results demonstrate that our intuition radically reshapes both topology and overall dataset dynamics, underscoring the need for this new GDB-GML architecture to produce better models and unlock the full potential of graph-based data analysis.
\end{abstract}

\begin{IEEEkeywords}
Graph Machine Learning, Graph Databases, Knowledge Graphs, Neo4j, Transitive Relationships, Knowledge Graph Reasoning, Knowledge Graph Completion.
\end{IEEEkeywords}

\section{Introduction}
Traditional machine learning (ML) models struggle with complex, heterogeneous and interconnected datasets, as they typically work on independent data points, missing relational interactions that are crucial to capture underlying patterns. 
Moreover, datasets exist in various forms and are constantly growing in both size and variety, demanding flexible database management systems (DBMSs) that efficiently handle  relationships, something that traditional one cannot accomplish\cite{b5}. These problems are especially evident in fields such as social networks and biology, where connections play an important role and cannot be properly modeled and analyzed\cite{b6}.

The increasing availability of complex datasets, the inefficiencies of traditional DBMSs and the limitations of ML techniques have raised the growth of Graph Machine Learning (GML) with NoSQL Graph Databases (GDBs). In particular, GML represents data as graphs, where nodes are entities and edges are relationships. This structure naturally capture real-world dynamics, by using techniques designed to learn from and reason about graph-structured data, 
like: node classification, edge prediction and graph classification\cite{b55}. 
Among NoSQL DBMS, GDBs raise due to their suitability for Big Data and ML applications that involve highly interconnected data\cite{b10}.
They also have a graph-based schema, offering an efficient representation and processing of complex relationships in GML tasks. 

Even if the union of these two technologies represents a rapidly evolving field, with numerous advantages and promising results\cite{b52}, the current GDB-GML architecture has significant limitations that compromise results' accuracy. In fact, the performed analysis are not accurate, as they rely only on a partial view of the information that the datasets actually contain. Since graphs consist of nodes and relationships with different semantic, the high heterogeneity of these elements makes some information not explicated. As a result, a dataset that may appear fragmented, incomplete, or lacking in content at first, could actually contain a substantial amount of not considered implicit knowledge. Since these incomplete datasets are used as input for GML systems, models analyze only a subset of relationships, reducing the accuracy of the results and missing the opportunity to reveal hidden insights.

For example, this is evident in GML fraud detection systems, where intermediaries help money launderers with multi-step transactions, and in recommendation engines, where missing implicit connections make it difficult to retrieve deep relationships, leading to less accurate suggestions.

Our key intuition is that, since GDB-GML are Knowledge Graph-based systems (i.e. graphs where nodes and relationships can be semantically different), their architecture  lacks a built-in mechanism for Knowledge Completion (KC), which is an overlooked critical aspect in the current workflow. In fact, KC is a technique designed to infer and fill in implicit relationships within KGs\cite{b34}, so its integration can completely reshapes the original graph structures, generating richer and more informative datasets that will be given as input for GML models. This integration significantly increases both the quantity and quality of the available data, leading to: more accurate real-world representations, enhanced model performance and improvements in GML downstream tasks, by improving node embeddings and centrality measures.

In particular, we introduce scalable transitive relationships, which are links that not only connect pairs of nodes but also propagate information across multiple entities. They are prevalent in nature, such as in genetic mutations, where the genes transmission decreases with family degree. However, despite their relevance, they are not considered in current GDB-GML applications since are in the KC domain. Their integration in a dataset can completely modify the network's topology, resulting in a very different input given to GML models. To describe these relationships, decay factors and aggregation functions are also introduced to model the attenuation of influence over distance or time.

In conclusion, this paper presents a novel GDB-GML systems architecture that uncovers hidden transitive relationships, preventing: incomplete data, misleading metrics and inaccurate models, thereby improving ML feature vector data.

\section{State of Art of NoSQL Graph Databases Analysis in Machine Learning} 
GDBs and GML models share the same way to represent data as a network of interconnected nodes (entities) and edges (relationships), with each of them having key-value pair properties\cite{b12}. 
In this context, Neo4j\footnote{neo4j.com} is the most widely used DBMS in GDB-GML applications, employing a data representation model based on KGs. It also provides libraries, such as Graph Data Science (GDS)
, that enable the implementation of GML algorithms, such as node embeddings technique like Node2Vec and  Graph Neural Networks (GNNs). 

The atomic unit in a KG is a triple, which consists of: a labelled head node $H$, a labelled tail node $T$ and a directed labelled edge that goes from the head to the tail\cite{b32}.
Formally, a KG is represented by the tuple  \( KG = (V, E, R) \), where:
\begin{itemize}
    \item \( V \) is the set of entities (nodes),
    \item \( E \subseteq V \times V \) is the set of directed edges,
    \item \( R \) is a set of relationship types that define the edges in \( E \).
\end{itemize}

Successful implementations of GML with Neo4j span across various sectors, like social networks, where Facebook and LinkedIn model user relationships for friend recommendations and advertising.  E-commerce, for product catalogs recommendation engines and 
in healthcare, as they model: patient relationships, medical histories and treatment pathways for better clinical decision-making processes\cite{b18}. 

However, results accuracy remain a crucial aspect, especially for massive datasets, where performance can be impacted by schema design, node embeddings and query strategies\cite{b21}.
In all these implementations, GDB-GML analysis started with Knowledge Fusion (KF), which involves integrating data from heterogeneous sources to create a semantically rich knowledge representation\cite{b42}, creating meaningful KGs. 
After creating the KG, different Neo4j libraries are directly used to perform Knowledge Reasoning (KR), that aims to predict new knowledge from existing information by calculating: communities, node embeddings and centrality measures as model features\cite{b45}. 

The most relevant error in all these works is the direct transition from KF to KR, which is a serious mistake when analyzing KG-based systems. In fact, while KF and KR are being explored and used in various applications, the lack of an integrated KC phase represents a significant limitation\cite{b48}.
In particular, the current GDB-GML architectures start with the KF phase, whose output, along with graph metrics and centrality measures, serves as the input for KR, 
where node embeddings are calculated and used in GML models.

\begin{figure}[ht!]
    \centering
    \includegraphics[width=1\linewidth]{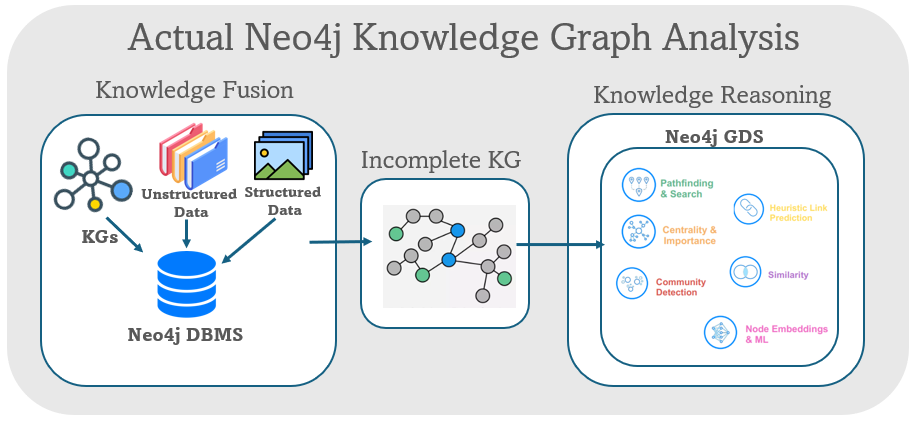}
    \caption{Actual Neo4j Analytics Flow.}
    \label{fig:actual_neo4j_flow}
\end{figure}

KC omission prevents the identification of implicit relationships that could drastically modify the graph's topology, resulting in wrong GML outcomes, as the reasoning rely on incomplete datasets and graph metrics that may hide important elements like: communities, relationships and central nodes. In particular, inaccurate graph metrics 
 leads to 
wrong embeddings, causing poor performance in ML downstream. 
For example, in healthcare, this problem impacts drug discovery and personalized medicine, while in finance,  leads to limited risk assessments\cite{b62}.

Only few recent works in literature analyze something similar, like \cite{b59} that highlights the potential of contextual Large Language Models (LLMs) like GPT-2 in KGs. 
However, their integration was not mainly related to KC and remains an open research problem, with challenges including computational cost and model interpretability. 

In conclusion, while the synergy between graph databases and GML holds immense potential for handling complex related data, the field remains relatively new, with some important workflow problems that limits the correctness of results.

\section{Improving Graph Machine Learning with Knowledge Completion on Graph Databases}
The integration of expert knowledge and domain-specific ontologies in GDB-GML workflow is crucial for developing more accurate and reliable GML system. This effort will unlock the full potential of NoSQL GDBs in providing comprehensive and insightful analyses for GML. 

By using Neo4j, which is the most important GDB, our idea is to change current GDB-GML systems' architecture (Fig.\ref{fig:actual_neo4j_flow}) using the correct KG-based analytics flow, that should be developed through three fundamental phases:
\begin{enumerate}
    \item Knowledge Fusion (KF), which is 
    the process of integrating and unifying knowledge from multiple sources to create a coherent KG, with the aim of integrate Big Data and merge existing Knowledge Graphs;
    \item Knowledge Completion (KC), which focuses on filling in missing entities and relationships that are implicit during KF process. 
    \item Knowledge Reasoning (KR), that involves inferring new facts from the complete KG with different methods like Expert Knowledge Integration (EKI) and ML models\cite{b35}.
\end{enumerate}

\begin{figure}[ht!]
    \centering
    \includegraphics[width=0.9\linewidth]{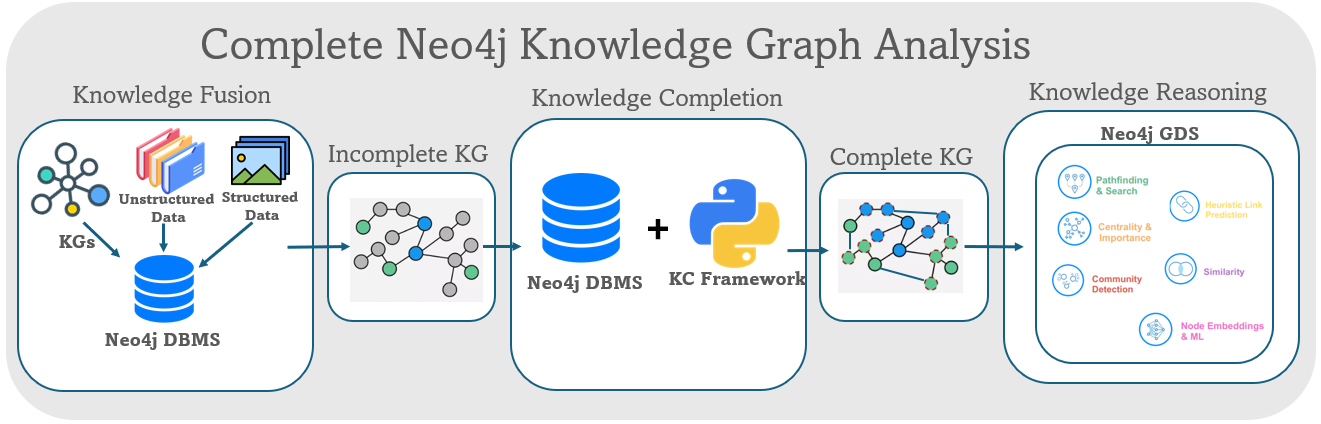}
    \caption{Fixed Neo4j Analytics Flow.}
    \label{fig:fixed_workflow}
\end{figure}

Our architecture consists of the KC phase introduction after KF and before KR, integrating all missing implicit edges related to a specific domain of knowledge, completing the graph's structure and recalculating its metrics. GML is then performed on the enriched graph, demonstrating how the integration of this intermediate step can significantly change reasoning features (Fig.\ref{fig:fixed_workflow}).
In particular, when integrating KC into a GML pipeline, GML models will ultimately learn and make inferences from richer data as:
\begin{itemize}
    \item Newly added edges or node attributes from KC become part of the GML feature space, as the number and nature of neighbors and paths feeding into ML algorithms shifts drastically with direct implications on node embeddings;
    \item KC reshapes topology and metrics, as GML pipelines also rely on graph metrics as input features or as a mechanism for subgraph selection;
\end{itemize}

Knowledge domains formed by transitive relationships will be considered for the experiments. This property is part of the so called Commonsense Logic, which refers to a form of reasoning related about general knowledge and assumptions about the world that are intuitive for humans. The current architecture ignores this concept, resulting in graphs that are completely different from real dynamics.

In mathematics, a relation $R$ on a set $A$ is called transitive if, for any elements $a$, $b$ and $c$ in \( A \), the following holds:
\[
a R b \text{ and } b R c \Rightarrow a R c
\]
In other words, if \( a \) is related to \( b \) and \( b \) is related to \( c \), then \( a \) must also be related to \( c \).
In KGs domain, a relationship \( r \in R \) is said to be \textit{transitive} if, for any entities \( x, y, z \in V \), the following holds:

\begin{equation}
(x \xrightarrow{r} y) \land (y \xrightarrow{r} z) \Rightarrow (x \xrightarrow{r} z)
\end{equation}

This means that if there exists a relationship \( r \) from \( x \) to \( y \) and from \( y \) to \( z \), then a direct relationship \( r \) must also exist from \( x \) to \( z \). 

Currently, only Translation-based embedding models are used to represent them, as translations in the vector space, allowing to recover partial transitive relationships only through a probabilistic approach and not with a deterministic model. Moreover, since it is performed in KR phase, once the embedding is created, the KG cannot be modified without needing a new one, increasing computational complexity\cite{b72}.

The core idea is to create a deterministic model that overcomes these limitations by discovering hidden relationships before KR. At the beginning, it assigns at each relationship a "transitivity level" indicator, which serves as a way to capture its ability to propagates over nodes. In particular, this transitivity level has a binary value representing non-transitive and transitive relationships respectively. For relationships marked as transitive, the model then enables propagation across different nodes, dynamically expanding the connectivity on these inferred relationships based on a decay function.

Formally, let \( KG = (V, E, R) \) be a KG as previously defined.
Each edge \( e_{ij} \in E \) between nodes \( v_i, v_j \in V \) is associated with a relationship type \( r \in R \). Relationships can be distinguished as either transitive or non-transitive, characterized by the indicator \( T(e_{ij}, r) \), which represents the deterministically inferred ability of the relationship to extend beyond direct links.

The transitivity indicator \( T(e_{ij}, r) \) is defined as:
\[
T(e_{ij}, r) \in \{0, 1\}
\]
where:
\begin{itemize}
    \item \( T(e_{ij}, r) = 0 \) indicates a \emph{non-transitive} relationship, meaning that the relationship \( r \) between \( v_i \) and \( v_j \) does not propagate to other nodes.
    \item \( T(e_{ij}, r) = 1 \) indicates a \emph{transitive} relationship, allowing \( r \) to propagate along paths, forming indirect connections in the graph.
\end{itemize}

For transitive relationships where \( T(e_{ij}, r) = 1 \), we enable propagation along a path of nodes. Given a path \( P = \{ e_{ij}, e_{jk}, \dots, e_{lm} \} \subset E \), where each edge \( e_{xy} \) along \( P \) is of type \( r \) with \( T(e_{xy}, r) = 1 \), we infer an indirect relationship of type \( r \) from \( v_i \) to \( v_m \). 

We also introduce the concept of scalability in transitive relationships, which can be described as the tendency of the link to weaken progressively as it propagates between nodes and along descendants,  useful for the analysis of genetic anomalies or hereditary diseases.

A transitive relationship \( r \) is said to be \textit{scalable} if there exists a function  

\[
S(x, z, r): \mathbb{N} \rightarrow [0, 1]
\]

such that assigns a value called strength at the transitive relationship between two nodes \( x \) and \( z \) depending on all the paths with the same relation $r$ connecting them. Each path is denoted as  

\[
p \in P(x, z, r),
\]

where \( P(x, z, r) \) represents the set of all paths between \( x \) and \( z \) composed exclusively of relationships of type \( r \).  

The strength of the relationship along a specific path \( p \) depends explicitly on the number of hops \( h(p) \) in the path. The strength for a specific path \( p \) is defined as:  

\[
S_p(x, z, r) = f(h(p)),
\]

where: \( x \) and \( z \) are nodes in the graph;  \( r \) represents the considered transitive relationship; \( h(p) \) denotes the length (number of hops) of the path \( p \); \( f(h) \) is a  decreasing function modeling the attenuation of strength as the number of hops increases.  

The overall transitive strength \( S(x, z, r) \) between \( x \) and \( z \) is obtained by aggregating the strengths of all valid paths using a function \( \mathcal{A} \), formally:  

\[
S(x, z, r) = \mathcal{A}\big( \{ S_p(x, z, r) \mid p \in P(x, z, r) \} \big),
\]

where:  
\begin{itemize}
  \item \( \mathcal{A} \) is an aggregation function (e.g., maximum, average, or sum) that combines the strengths of individual paths based on the context;  
  \item \( S_p(x, z, r) \) denotes the strength computed for a specific path \( p \) connecting \( x \) and \( z \);  
  \item \( P(x, z, r) \) represents the set of all paths between \( x \) and \( z \) restricted to relationships of type \( r \).  
\end{itemize}  

Finally, the relationship propagates if the overall strength satisfies:  

\[
S(x, z, r) > \tau,
\]

where \( \tau \) is a context-dependent threshold.

Initially, all paths \( P(x, z, r_t) \) between pairs of nodes \( x \) and \( z \) are identified, where each path consists exclusively of edges labeled with the transitive relationship \( r_t \). For each path \( p \), the partial strength \( S_p(x, z, r_t) \) is computed using a decay function \( f(h(p)) \), where \( h(p) \) denotes the number of hops in the path \( p \). The overall strength \( S(x, z, r_t) \) is then obtained by aggregating the strengths of all paths using a function \( \mathcal{A} \). If the aggregated strength exceeds a predefined threshold \( \tau \), it is added to the KG with the associated strength.

In this study, we explore how the outcomes of common graph theory metrics change with our approach, considering natural directions of edges:
\begin{itemize}
    \item Degree Centrality, which calculates the total number of connections for each node;
    \item PageRank, that measures the importance of nodes  based on the quality and quantity of their connections;
\end{itemize}

By applying transitive inferences to relationships, we aim to demonstrate how centrality metrics are affected, revealing changes in node roles and influence within each graph that would change the input of GML models.

\section{Experiments}
To highlight the impact of KC and transitive scalable relationships, we evaluate the changes on traditional graph metrics and node properties, which are used as basis for GML pipelines. Experiments were conducted using Neo4j on two distinct datasets. On these, we made KC and then calculated different centrality measures, comparing the results between our proposed system with the normal Neo4j-GML one. 

The first dataset represents the hierarchical administrative structure of the Roman Empire during the reign of Diocletian, constructed to illustrate the concepts of \textit{Divide et Impera}. 
It describes a hierarchical system of governance in which: at the top was the Emperor, followed by four Prefectures, subdivided into Dioceses that were further divided into Provinces\cite{b36}. 

Rome served as the ultimate source of authority, with every decision and command originating from the Emperor and propagating directly to the region of interest or through the hierarchical structure.

\begin{figure}[ht!]
    \centering
    \includegraphics[width=0.7\linewidth]{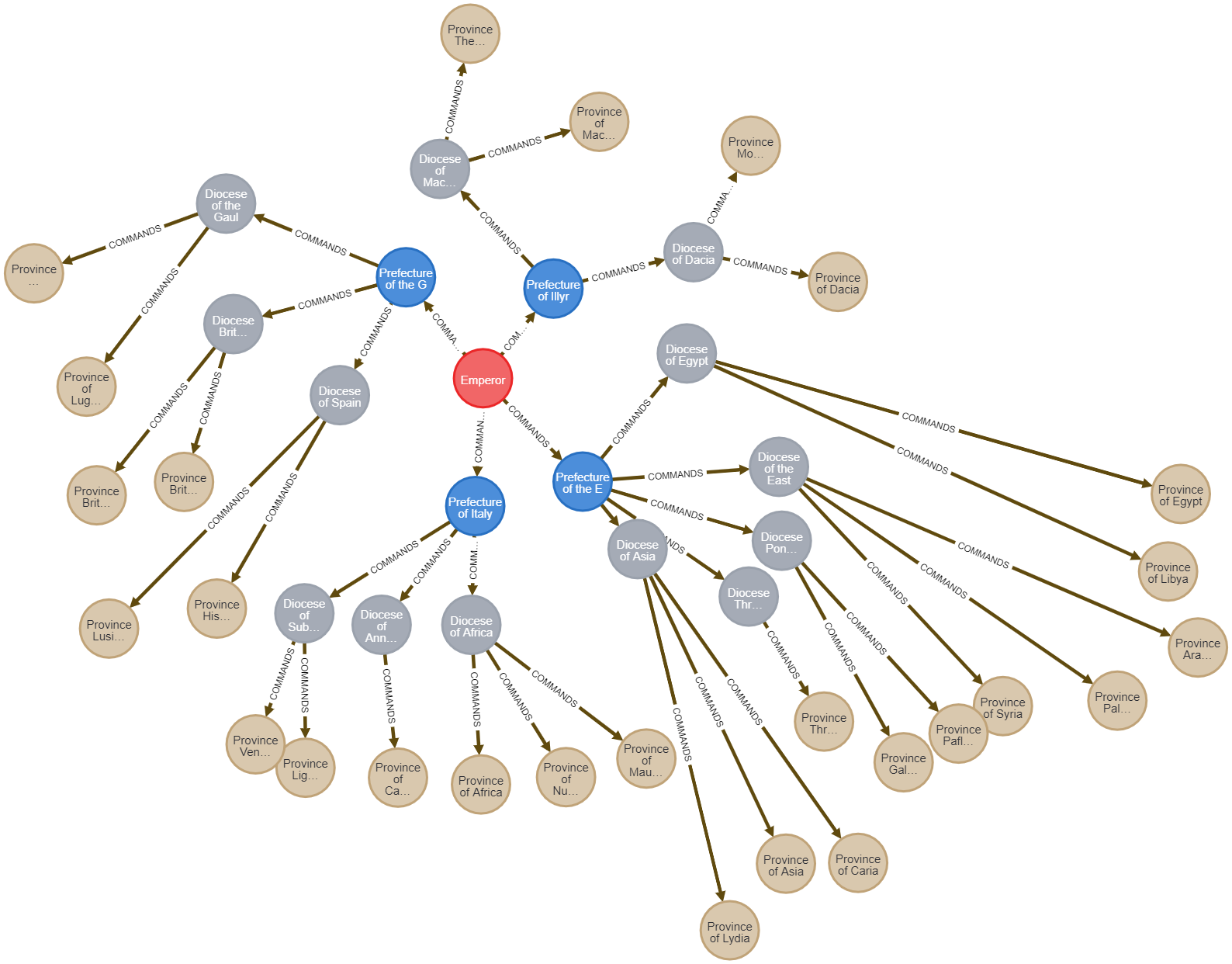}
    \caption{Roman Empire Initial Dataset.}
    \label{fig:roman_empire_initial_graph}
\end{figure}

\begin{figure}[ht!]
    \centering
    \includegraphics[width=0.75\linewidth]{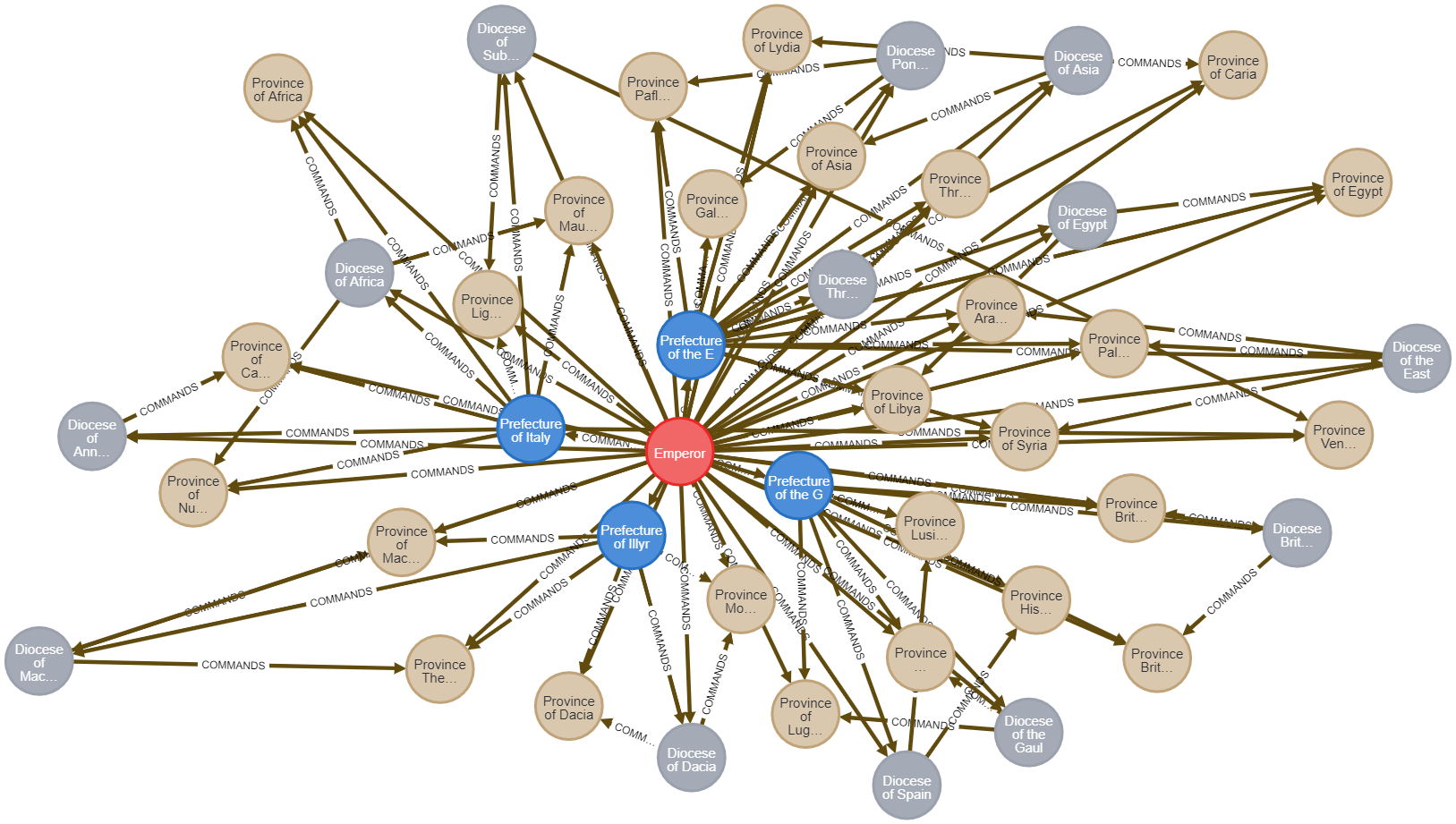}
    \caption{Roman Empire with Knowledge Completion.}
    \label{fig:roman_empire_kc}
\end{figure}

Formally, the administrative structure can be modeled as a KG, with \( R = \{\text{COMMANDS}\} \) is the set of relationships, where \(\text{COMMANDS}(v_i, v_j)\) denotes that \( v_i \) has authority over \( v_j \).
The initial KG graph only captures limited direct relationships (Fig. \ref{fig:roman_empire_initial_graph}).
Applying KC with the transitivity 
transforms it, filling in 67 indirect links from the initial 47 and making it almost fully connected (Fig. \ref{fig:roman_empire_kc}).

The initial graph analysis revealed a hierarchical distribution of centrality. However, applying KC with transitive relations reveals that the Emperor and the Prefectures hold an even more central position than initially indicated by the dataset. At the opposite, some Dioceses have become less significant as new paths emerged. This is a logical outcome, but it was not evident and explicated in the original dataset.

\begin{figure}[ht!]
    \centering
    \includegraphics[width=1\linewidth]{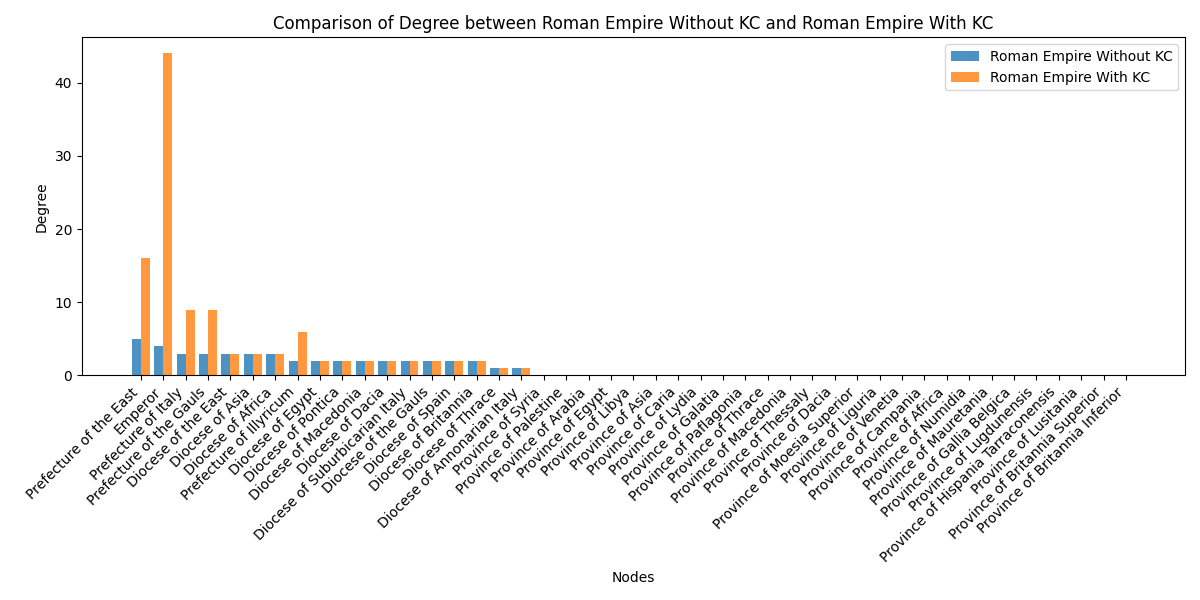}
    \caption{Roman Empire Degree Centrality Comparison}
    \label{fig:roman_empire_degree_centrality_comparison}
\end{figure}

In particular, the first experiment, which is focused on degree centrality (Fig. \ref{fig:roman_empire_degree_centrality_comparison}), reveals that adding implicit relationships has generally increased this value for most nodes, especially for those that gained additional direct connections. Finally, comparing the results of the two workflows, it is clear that the role of the Emperor in the network is now the most important with respect to the standard analysis.
\begin{figure}[ht!]
    \centering
    \includegraphics[width=0.65\linewidth]{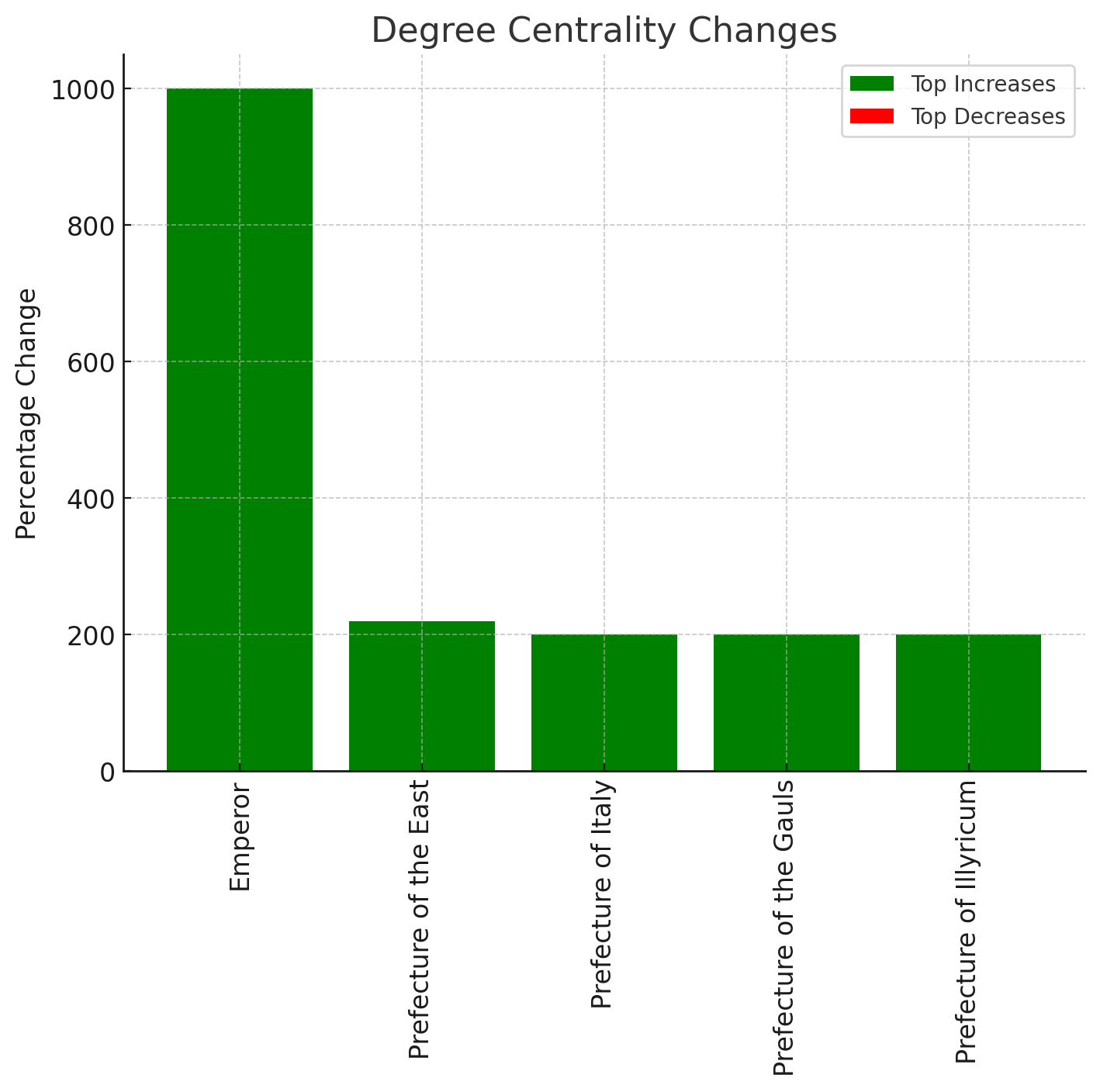}
    \caption{Most important Degree Centrality Changes in percentage.}
    \label{fig:enter-label}
\end{figure}

In particular, the Emperor experienced an increase of 1000\%\, reflecting the addition of numerous connections that make this node a central hub. Administrative regions like: the Prefecture of the East, Prefecture of Italy and of the Gauls also saw notable increases of 220\%\ and 200\%\, indicating the establishment of new relationships that increased their importance in the network. At the opposite, many provincial nodes retained a degree centrality of zero, showing that KC primarily affected higher-level administrative structures rather than peripheral regions. 

Similarly, in the second experiment, in which we considered Pagerank algorithm, after adding explicit relationships the PageRank values of several nodes increased, reflecting their increased influence within the network. 
This also reduces the dependency on certain intermediary nodes that were previously essential for connecting parts of the network, thereby redistributing influences. As a result, nodes with new direct relationships become more important, while some central nodes in the previous structure may see a decrease as influence spreads across a more interconnected network. 
\begin{figure}[ht!]
    \centering
    \includegraphics[width=1\linewidth]{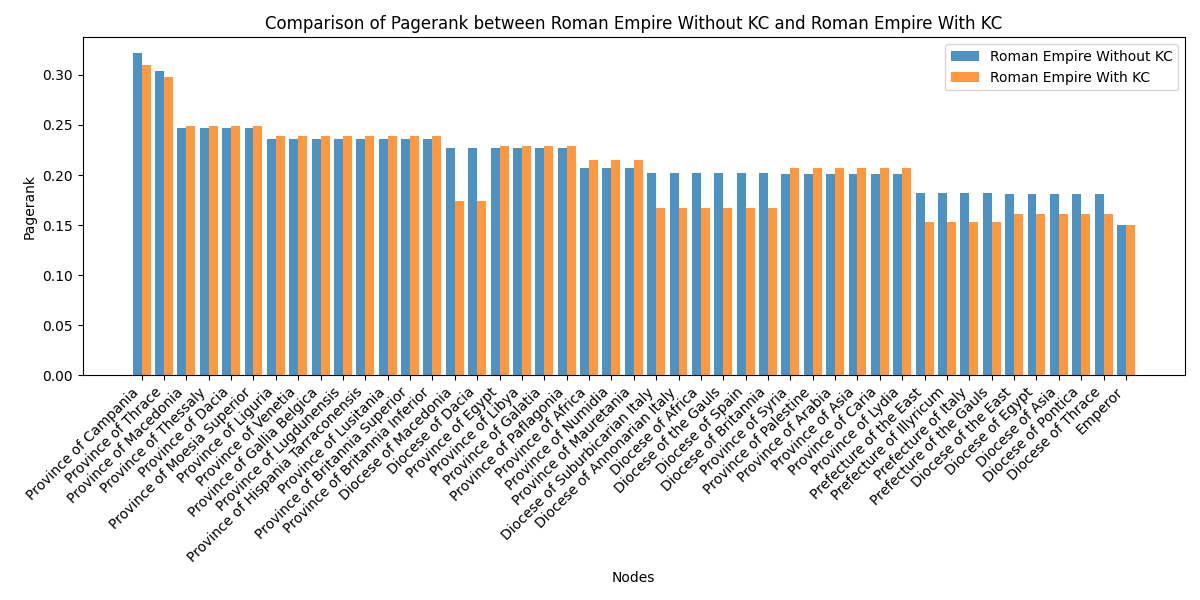}
    \caption{Roman Empire PageRank Comparison.}
    \label{fig:roman_empire_pagerank_comparison}
\end{figure}
\begin{figure}[ht!]
    \centering
    \includegraphics[width=0.65\linewidth]{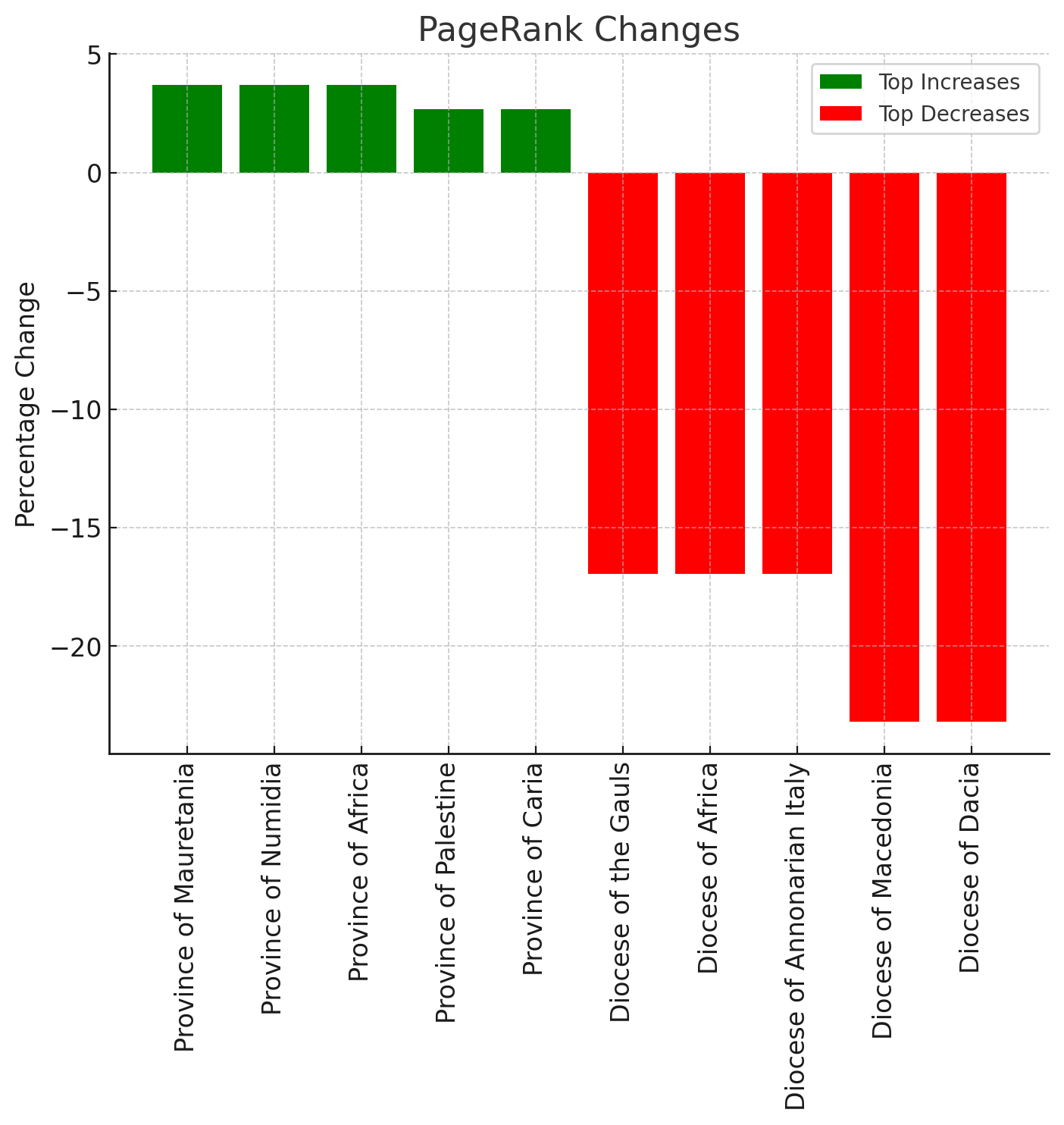}
    \caption{Most Relevant Pagerank results in percentage.}
    \label{fig:most_important_pagerank_changes}
\end{figure}

The analysis of PageRank with KC shows many shifts in node importance. Provinces like: Mauretania, Numidia and Africa experienced the most significant increases, with their values rising by approximately 3.69\%\ . These changes indicate that the new workflow highlighted their roles as influential nodes within the network by adding new connections. Other provinces, such as Palestine and Caria, saw smaller but still meaningful increases of around 2.67\%\ .
At the opposite, administrative regions such as the Diocese of Macedonia and Diocese of Dacia experienced substantial decreases in PageRank, with drops of 23.2\%\ . 
These reductions suggest that their relative influence in the network decreased as the added edges redistributed importance among nodes.


The main focus on the second dataset was on Kinship, which is linked to shared genetic material and crucial in genealogy. Understanding it, involves quantifying shared DNA percentage and its change across generations. 
The amount of shared DNA typically becomes negligible after approximately seven generations (or "hops")\cite{b69}, which are the generational steps separating individuals in a family tree.
GML algorithms can effectively process this graph-structured data to identify anomalies and predict relationships  but they also need a very accurate structure in order to perform well\cite{b71}.

The genealogical tree of the UK Royal Family of House of Saxe-Coburg and Gotha / House of Windsor was then analyzed with our innovative framework. Let be \( KG = (V, E, R) \) a KG, where \( R = \{RELATIVE-OF\} \)  and  \(RELATIVE-OF(v_i, v_j) \) denotes that \( v_i \) is a relative of \( v_j \) with a certain shared DNA.
Since the relationship is transitive, it propagates along paths following \( f(d(x, z)) \), which is the decay function the represent the shared DNA quantity in percentage and it is defined as:
    \[
    f(d) = \left(\frac{1}{2}\right)^d
    \]

Considering the aggregation function $A$ as sum of DNA fragments,  a threshold \( \tau= \frac{1}{2}^7  \) is applied such that relationships with \( S(x, z, r) < \tau \) are excluded from the graph. 


\begin{figure}[ht!]
    \centering
    \includegraphics[width=0.6\linewidth]{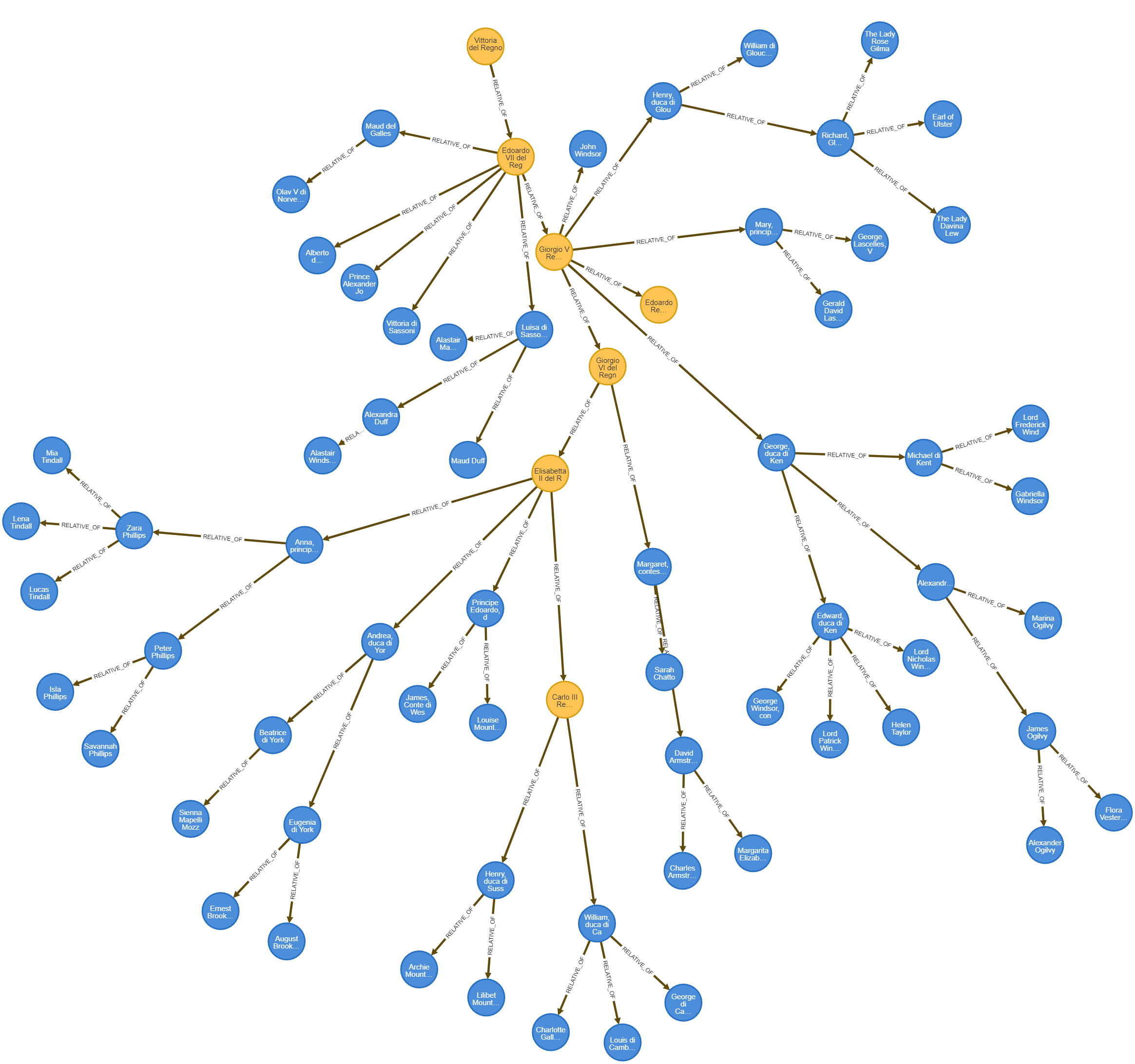}
    \caption{Royal Family Tree before KC (yellow nodes are Kings or Queens).}
    \label{fig:enter-label}
\end{figure}
\begin{figure}[ht!]
    \centering
    \includegraphics[width=0.6\linewidth]{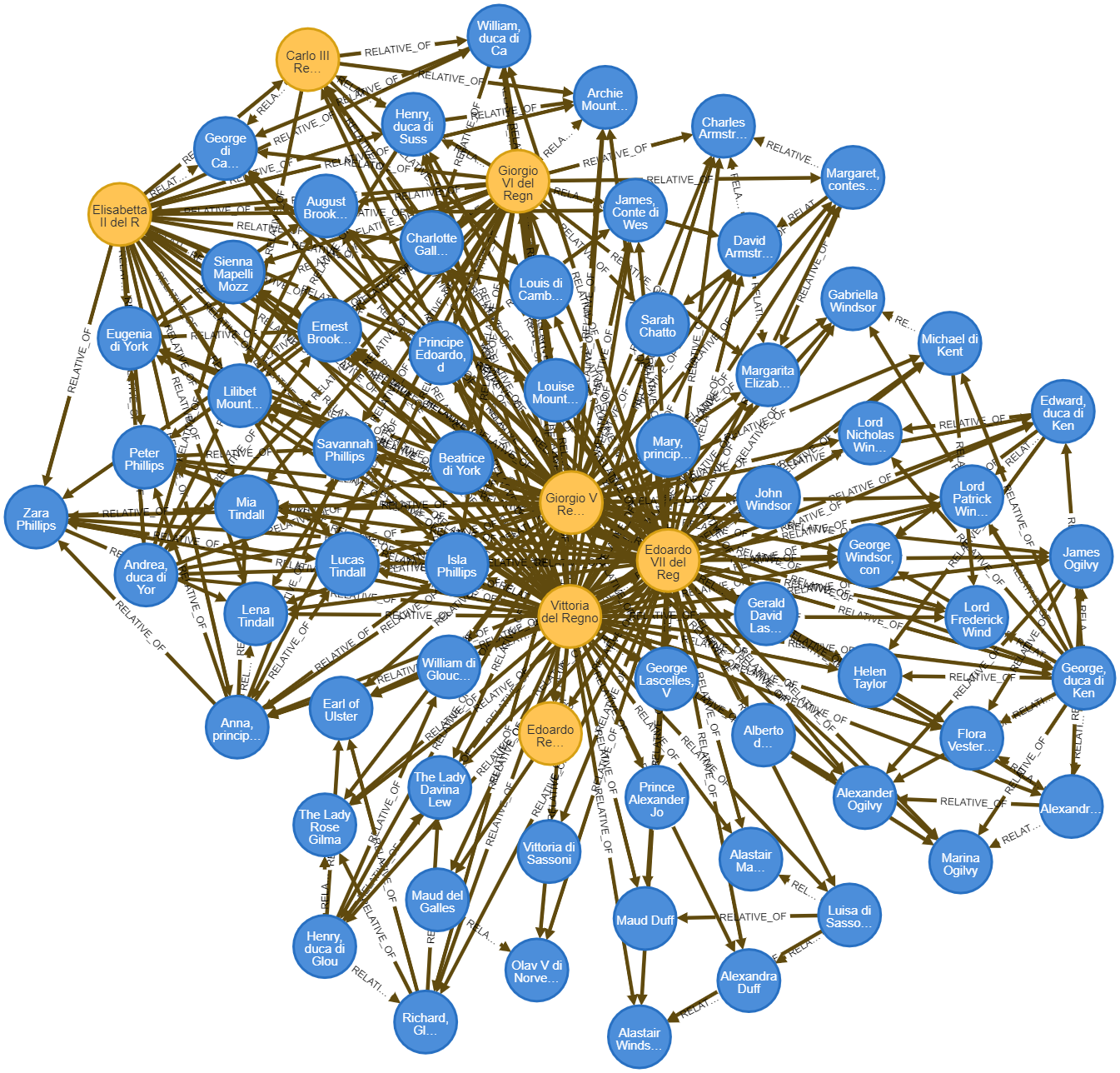}
    \caption{Royal Family Tree after KC.}
    \label{fig:enter-label}
\end{figure}

With the KC, 598 previously missing relationships were added, allowing us to observe not only a simple tree but the true extent of shared DNA between different dynasties and to identify the most significant branches of the royal family with greater accuracy, offering deeper insights into the hereditary links that shaped royal bloodlines over time.

\begin{figure}[ht!]
    \centering
    \includegraphics[width=1\linewidth]{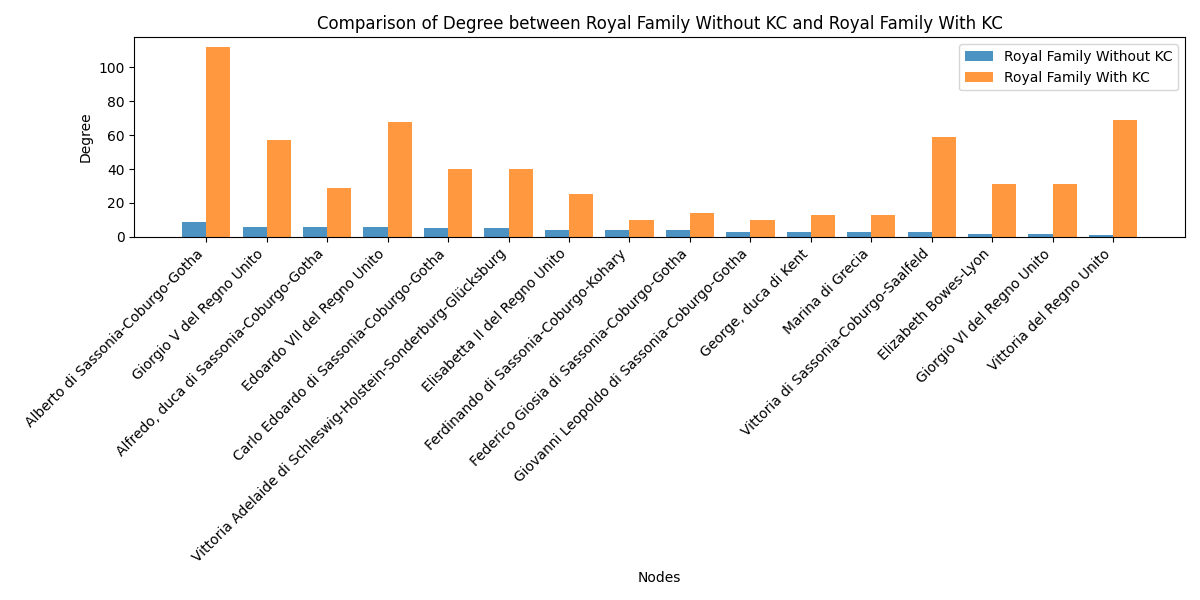}
    \caption{Royal Family's Degree Centrality before and after KC.}
    \label{fig:rf_degree}
\end{figure}

\begin{figure}[ht!]
    \centering
    \includegraphics[width=1\linewidth]{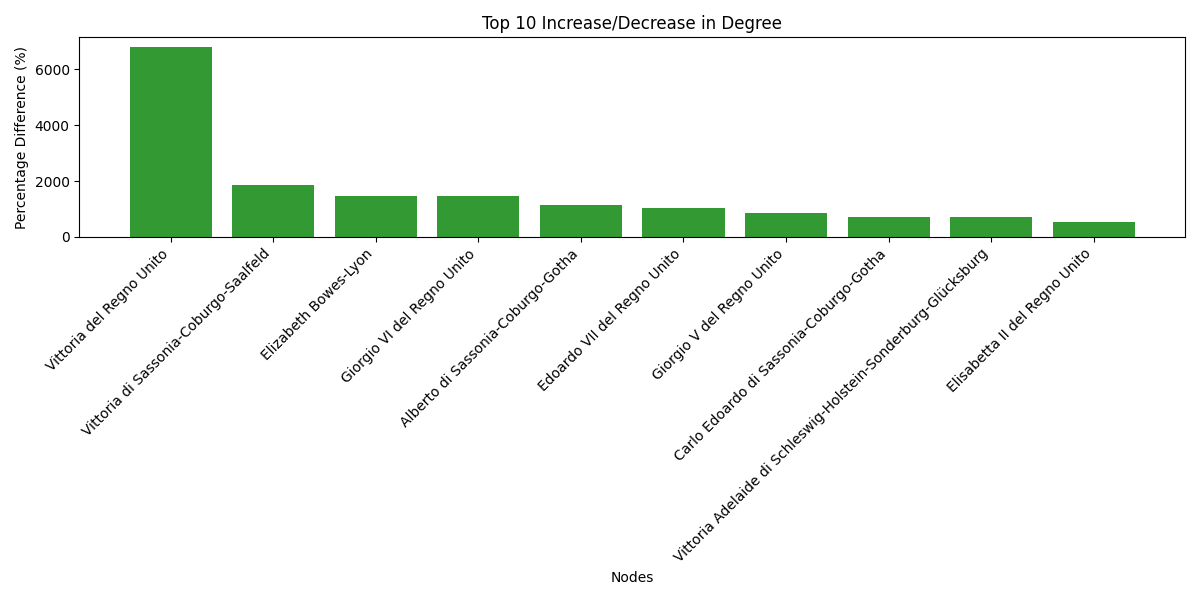}
    \caption{Royal Family's Degree Centrality Top Increase/Decrease.}
    \label{fig:rf_top_degree}
\end{figure}

The first experiment, which is the degree centrality, Albert of Saxe-Coburg-Gotha experienced an increase  from 9.0 to 112.0, a change of +1144\%\ while Queen Victoria experienced the highest percentage increase variation  of +6000\%\ (Fig. \ref{fig:rf_degree} and Fig. \ref{fig:rf_top_degree}), making them the most prolific ancestor in terms of genetic contribution. 

\begin{figure}[ht!]
    \centering
    \includegraphics[width=1\linewidth]{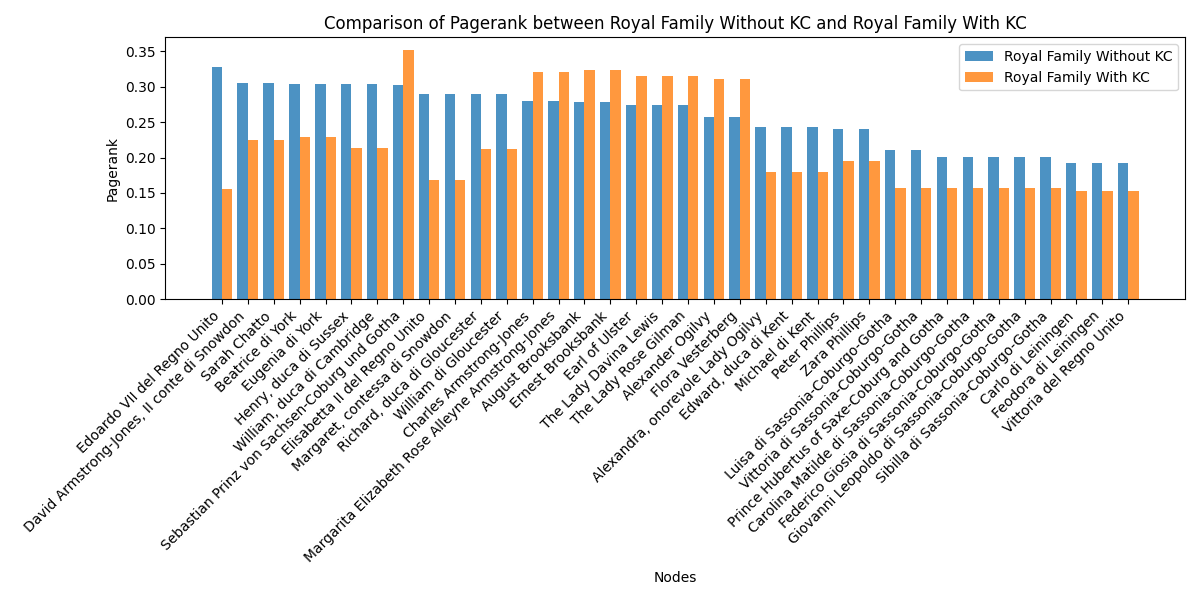}
    \caption{Royal Family's Pagerank before and after KC.}
    \label{fig:rf_pagerank}
\end{figure}

\begin{figure}[ht!]
    \centering
    \includegraphics[width=1\linewidth]{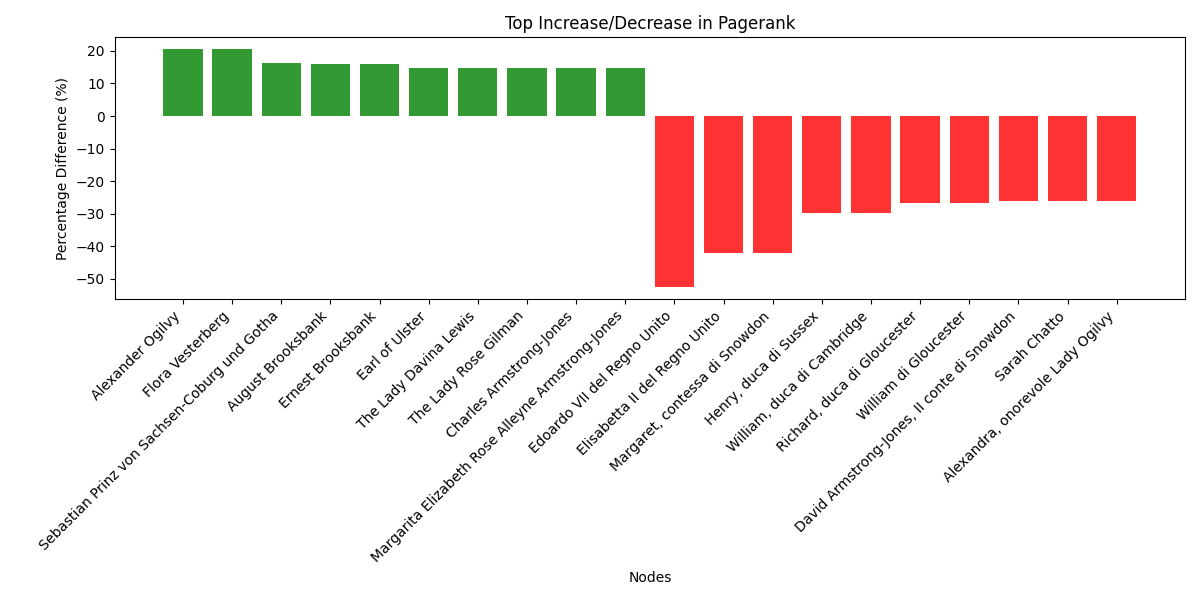}
    \caption{Royal Family's Top Pagerank Increase/Decrease.}
    \label{fig:rf_top_pagerank}
\end{figure}

Several individuals in the genealogical tree experienced notable increases in PageRank (Fig. \ref{fig:rf_pagerank}) after the application of transitivity. For instance, Flora Vesterberg and Alexander Ogilvy both saw a substantial 20.51\%\ increase (Fig. \ref{fig:rf_top_pagerank}) in their PageRank, rising from 0.2576 to 0.3105. This significant rise indicates that their influence within the lineage expanded through inherited mutations and recombination, amplifying their role in the transmission of genetic material. 

\section{Conclusions}
In conclusion, our proposed system, enhanced with scalable transitive relationships in KC, demonstrates remarkable potential for future applications in GML, addressing critical limitations of traditional approaches, particularly in their ability to  support advanced graph-based reasoning and learning tasks.
By explicating implicit relationships, the graph becomes far richer and more interconnected, unlocking the correct usage of graph-theory metrics in a meaningful way. 
This enriched representation enables GML models to learn from a dataset that more accurately reflects real-world complexities, leading to improved performance.

If it is true that this process raises concerns about storage requirements and scalability, since lots of new inferred edges are added, it is also true that Neo4j's robust big data capabilities mitigate these challenges to obtain  benefits in terms of represented knowledge.
Future research will include a thorough scalability and performance evaluation and the integration of this workflow in GNNs, comparing results with and without the enriched KGs for training and testing more robust models.

The usage of this new workflow with scalable transitive relationships represents a pivotal advancement in GDB-GML applications, unlocking new possibilities for reasoning and GML analysis. By addressing the identified challenges, future iterations of this workflow can establish a robust foundation for applications in the ML domain. Furthermore, existing works that currently do not employ these architectures could be revisited to compare their results and generated datasets, potentially revealing undiscovered insights. 

\section{Acknowledgment}
Rosario Napoli is a PhD student enrolled in the National PhD
in AI, XL cycle, course on Health and
life sciences.
This work has been partially funded by the Italian Ministry of Health, Piano Operativo Salute (POS) trajectory 2 ``eHealth, diagnostica avanzata, medical device e mini invasività'' through the project ``Rete eHealth: AI e strumenti ICT Innovativi orientati alla Diagnostica Digitale (RAIDD)'' (CUP J43C22000380001) and ``SEcurity and RIghts in the CyberSpace (SERICS)'' project (PE00000014), under the MUR National Recovery and Resilience Plan funded by the European Union - NextGenerationEU (CUP: D43C22003050001) and the Italian Ministry of University and Research (MUR) ``Research projects of National Interest (PRIN-PNRR)'' through the project “Cloud Continuum aimed at On-Demand Services in Smart Sustainable Environments” (CUP: J53D23015080001- IC: P2022YNBHP).
\bibliographystyle{unsrt}
\bibliography{bib}

\end{document}